%% file: main.tex
\documentclass[conference]{IEEEtran}
\IEEEoverridecommandlockouts
\usepackage{cite}
\usepackage{amsmath,amssymb,amsfonts}
\usepackage{algorithmic}
\usepackage{graphicx}
\usepackage{textcomp}
\usepackage{xcolor}
\usepackage{multirow}
\usepackage{authblk}
\usepackage{setspace}
\def\BibTeX{{\rm B\kern-.05em{\sc i\kern-.025em b}\kern-.08em
    T\kern-.1667em\lower.7ex\hbox{E}\kern-.125emX}}

\makeatletter
\def\@maketitle{%
  \newpage
  \null
  \vskip 2em%
  \begin{center}%
  \let \footnote \thanks
    {\LARGE \@title \par}%
    \vskip 1.5em%
    {\large
      \lineskip .5em%
      \begin{tabular}[t]{c}%
        \baselineskip=12pt
        \@author
      \end{tabular}\par}%
    \vskip 1em%
    {\large \@date}%
  \end{center}%
  \par
  \vskip 1.5em}
\makeatother

\begin{document}

\title{\huge Audio-visual target speaker enhancement on multi-talker environment using event-driven cameras\\
\thanks{This work was supported by the European Union’s Horizon2020 project ECOMODE (grant No 644096).}
}
\vspace{8 pt}

\author[1]{\normalsize Ander Arriandiaga}
\author[2]{\normalsize Giovanni Morrone}
\author[3]{\normalsize Luca Pasa}
\author[3]{\normalsize Leonardo Badino}
\author[1]{\normalsize Chiara Bartolozzi}

\affil[1]{\small iCub Facility, Istituto Italiano di Tecnologia, Genova, Italy}
\affil[2]{\small Department of Engineering ”Enzo Ferrari”, University of Modena and Reggio Emilia, Modena, Italy}
\affil[3]{\small CTNSC, Istituto Italiano di Tecnologia, Ferrara, Italy}


\date{}
{\let\clearpage\relax%
\maketitle }

%
\begin{abstract}
We propose a method to address audio-visual target speaker enhancement in multi-talker environments using event-driven cameras. 
State of the art audio-visual speech separation methods shows that crucial information is the movement of the facial landmarks related to speech production. However, all approaches proposed so far work offline, using frame-based video input, making it difficult to process an audio-visual signal with low latency, for online applications. In order to overcome this limitation, we propose the use of event-driven cameras and exploit compression, high temporal resolution and low latency, for low cost and low latency motion feature extraction, going towards online embedded audio-visual speech processing. 
We use the event-driven optical flow estimation of the facial landmarks as input to a stacked Bidirectional LSTM trained to predict an Ideal Amplitude Mask that is then used to filter the noisy audio, to obtain the audio signal of the target speaker.
The presented approach performs almost on par with the frame-based approach, with very low latency and computational cost.
\end{abstract}


\section{Introduction}
\label{sec:intro}
\input{introduction.tex}

\section{Methods}
\label{sec:methodology}
\input{methodology.tex}

\section{Experimental setup}
\label{sec:experiments}
\input{experimental_setup.tex}

\section{Results}
\label{sec:discussion}
\input{results.tex}

\section{Conclusions and future work}
\label{sec:conclusion}
\input{conclusions.tex}

\bibliographystyle{IEEEtran}
\bibliography{references}

\end{document}

%% file: introduction.tex
\begin{figure*}[th]
    \centering
    \includegraphics[width=18cm, keepaspectratio]{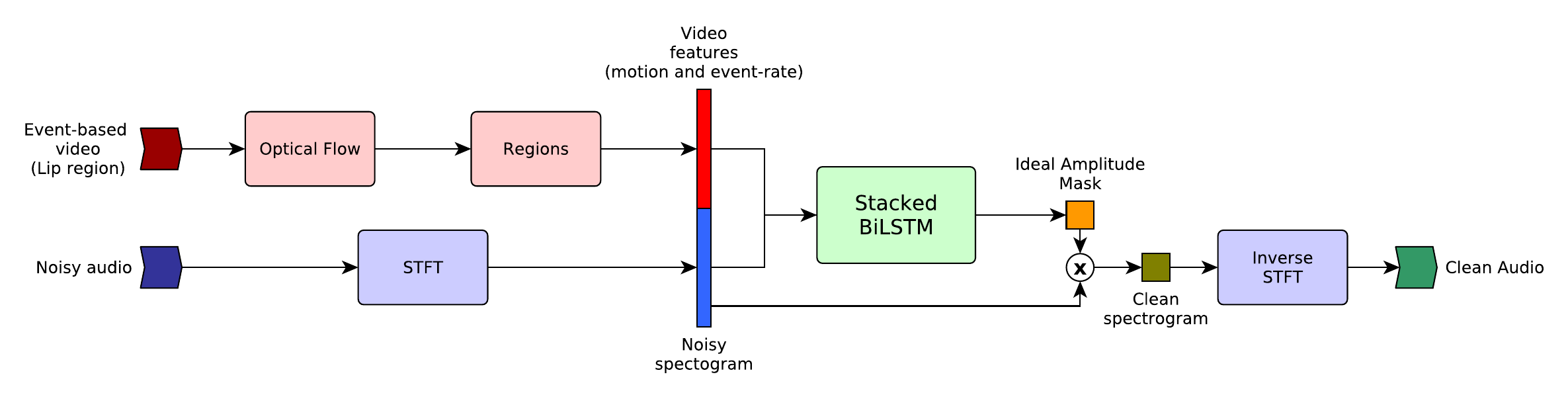}
    \caption{Audio-Visual Speech Separation pipeline.}
    \label{fig:pipeline}
\end{figure*}

The ability to disentangle and correctly recognise the speech of a single speaker among other speakers (the well known cocktail party effect~\cite{Cherry}) is paramount for effective speech interaction in unconstrained environments. As such, it is an extremely useful feature for any artificial device that relies on speech interaction such as robots and mobile devices. To this aim, it is crucial to devise efficient speaker enhancement techniques that rely on small datasets and low power sensing and computation. Humans solve this problem using complementary and redundant strategies such as physical sound source separation (thanks to stereo sound acquisition~\cite{Bronkhorst}) and using cues from lips motion~\cite{VisualNeuro}.

Artificial systems use single-channel audio signals as input to Long-Short Memory Networks (LSTM)~\cite{Isik2016SingleChannelMS, Kolbaek17, Deep_attractors, Tasnet} or dilated convolutional layers~\cite{Conv-TasNet} for speaker-independent enhancement. However, the number of speakers has to be known in advance, as well as the correspondence between the target speaker and the output clean speech. An alternative is to give as input to the model speaker dependent target features~\cite{zmolikova2017speaker, VoiceFilter}, using an LSTM-based speaker encoder to produce speaker-discriminative embeddings. However, this solution needs a reference utterance of the speaker and an additional trainable Deep Neural Network (DNN), making the speech separation performance conditioned on the performance of the speaker encoder network, and computationally heavy.

Inspired on the findings that viewing the target speaker's face improves the listener ability to track the speech~\cite{VisualNeuro}, methods that combine visual cues and speech processing achieved remarkably good results. They were based on residual networks (ResNet~\cite{ResNet}) pre-trained on a word-level lip-reading task \cite{Conversation,TimeDomainAV}, or based on a pre-trained face recognition model, in combination with 15 dilated convolutional layers~\cite{LookingToListen}. 
Such architectures, however, are computationally heavy and require heterogeneous and large audio-visual datasets for training.
An approach that allows to use smaller datasets (such as the GRID dataset~\cite{GRID}) is to rely on pre-trained models, with the use of images and corresponding optical flow as inputs to a pre-trained dual tower ResNet extracting visual features~\cite{Seeing}. 

If video features are extracted without using trainable methods, the neural networks are smaller and can be trained with smaller datasets without overfitting. Following this idea, \cite{Morrone} used face landmark movements as input visual features to a bidirectional LSTM that achieved good speaker-independent results on the GRID dataset. In this work, the use of landmark motion features rather than positional features turned out to be a key factor. Inspired by this finding, we propose to substitute the visual pipeline implemented with traditional frame-based sensors, face tracking and extraction of motion landmarks, with an equivalent pipeline, based on the use of a novel type of vision sensors -- the event-driven cameras (EDC) -- from which the extraction of motion is available at lower computational cost and latency. 
EDCs asynchronously measure the brightness change for each pixel, featuring a temporal resolution as high as $1\ \mu$s, extremely low latency and data compression (as only active pixels communicate data). With such an input, the audio-visual system can use the same temporal discretization of the auditory pipeline (10 ms), rather than the one of the visual pipeline (30 fps is the standard frame-rate of traditional sensors). 
Event-driven vision sensors have been widely used with good results for object tracking~\cite{ObjectTracking_Arren,ObjectTracking_Ryad},  detection~\cite{ObjectDetection} and segmentation~\cite{ObjectSegmentation}, and for gesture recognition~\cite{GestureRecognition}. Recently, they have been applied in the context of speech processing: vision-only speech recognition (i.e., lip-reading) on GRID exploited EDCs as input to a Deep Neural Network architecture~\cite{SpeechRecognition}; lip movements detected by an EDC were used to detect speech activity and enable an auditory-based voice activity detection
~\cite{Arman}, for embedded applications that require low computational cost. 
This work presents an audio-visual target speaker enhancement system on multi-talker environment using event-driven vision sensors that compute motion at lower latency and computational cost. Following~\cite{Morrone}, we propose a non trainable method to extract visual features combined with deep learning techniques. We use the GRID corpus in order to compare this approach with frame-based methods. To the best of our knowledge, this is the first work that presents an audio-visual target speaker enhancement system that uses event-driven cameras.

%% file: methodology.tex
This work is based on~\cite{Morrone} where audio and visual motion features were used as inputs of LSTM-based models to generate time-frequency masks. These masks are then applied to generate the clean spectrogram of the target speaker. 
Our contribution stands in the use of EDC for the acquisition of the visual signal and for the computation of the motion features of speech-production facial landmarks, based on the estimation of the normal optical flow from the events.
Fig.~\ref{fig:pipeline} shows the block diagram of the whole system we propose for audio-visual target speaker enhancement: the computed visual motion features relative to the target speaker are concatenated with audio features to train a Recurrent Neural Network (RNN) that estimates a time-frequency mask that, multiplied by the noisy spectrogram, separates the clean speech produced by the target speaker.

\subsection{Event-driven motion features extraction}
\label{ssec:eventdrivencamera}
EDCs output asynchronous events whenever a pixel detects changes in log intensity larger than a threshold. Each event has an associated timestamp, $t$, pixel position, $<x, y>$, and polarity (log intensity increase or decrease), $p$~\cite{ATIS}. The events are emitted with high temporal resolution and low latency, only when there is relative motion between the camera and the scene, increasing for fast moving objects and decreasing for less active scenarios, such as that of a target speaker talking in front of the camera. Fig.~\ref{fig:EV_face} shows the typical output in such scenario, where only the motion of the person and his/her mouth and eyes generate events, leading to a low amount of information to process and the possibility to have an always on front end visual acquisition and processing for audio-visual tasks. 
The different data structure and content from EDC require algorithms for the estimation of the optical flow, that can rely on the precise time of each event and the continuous observation of the events produced by contrast edges moving from one pixel to its neighbours. Even though the state of the art for EDC optical flow estimation is based on the use of deep learning~\cite{OF_SOA}, as the motion of lips and other facial landmarks are mostly perpendicular to the edge, we resorted to a temporally and computationally efficient algorithm for the estimation of the normal optical flow~\cite{OpticalFlow}.



%% file: experimental_setup.tex

\subsection{Dataset}
\label{ssec:dataset}
We focussed our analysis on a challenging and common scenario, where the quantity of available data and resources are limited, using the GRID dataset~\cite{GRID}. The dataset consists of $3$ seconds long audio and video recordings of 34 speakers pronouncing 1000 sentences in front of a frame-based camera and microphone. The camera records data at 25 frames per second, while the microphone data is recorded at 50kHz. We used a subset of the GRID corpus, consisting of 200 sentences from 33 speakers (one was discarded because the videos were not available). To test speech enhancement of a target speaker, the audio signals of two different speakers were mixed so that for each speaker there are 600 mixed-audio sample recordings.
From the total amount of samples, samples from 25 speakers were for training, from 4 speakers for validation and from the last 4 speakers for testing the model.
The videos were upscaled to 60 frames per second using video processing software to have more temporal information and avoid artefacts in the generation of events. The event-based data stream was generated by pointing the ATIS event-driven camera~\cite{ATIS} ($240\times304$ pixels with 8mm lens) towards a high definition LED monitor while the upscaled videos were played. Due to the low quality of the original videos ($360\times288$ pixels resolution) and in order to preserve the details in lip movements, we cropped the mouth area over $100\times50$ pixels from the event stream.
\begin{figure}[!t]
  \centering\includegraphics[width=0.8\linewidth]{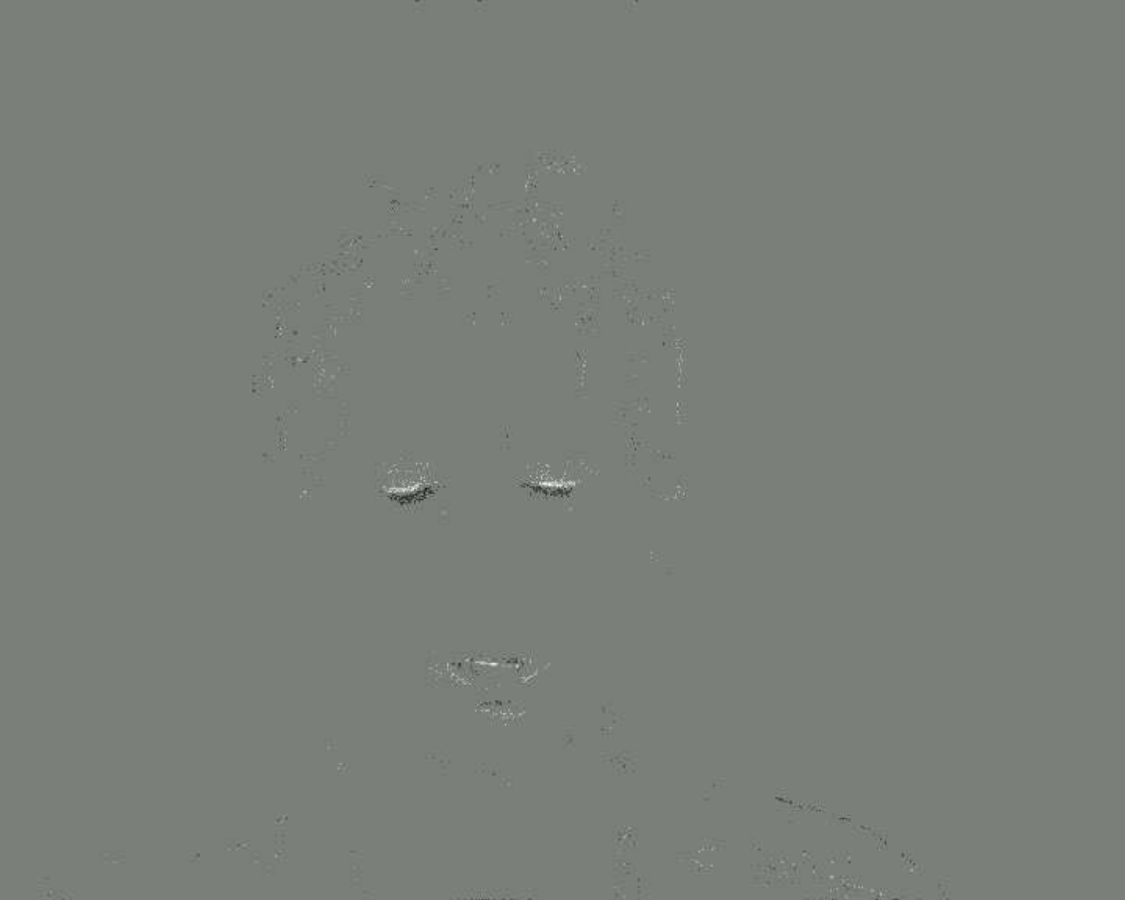}
  \caption{Snapshot of a person talking in front of an EDC camera}
  \label{fig:EV_face}
\end{figure}
\subsection{Model training}
\label{ssec:subhead}

\subsubsection{Audio pre- and post-processing}
\label{ssec:audioprocessing}
Following the state of the art in speech separation and enhancement, the audio original waveforms were pre-processed through Short Time Fourier Transform (STFT) applied over the over the audio waveforms resampled at 16kHz. STFT was applied using Fast Fourier Transform (FFT) size of 512, Hann window of length 25 ms, and hop length of 10 ms. The spectrogram \(|x|^p \) of each input audio sample was obtained performing power-law compression of the STFT magnitude with p = 0.3. Finally, the data was normalized per-speaker with 0 mean and 1 standard deviation. To reconstruct the clean audio, on the post-processing stage, the inverse STFT to the estimated clean spectrogram was applied using the phase of the noisy input signal.

\begin{figure}[!t]
  \centering\includegraphics[width=\linewidth]{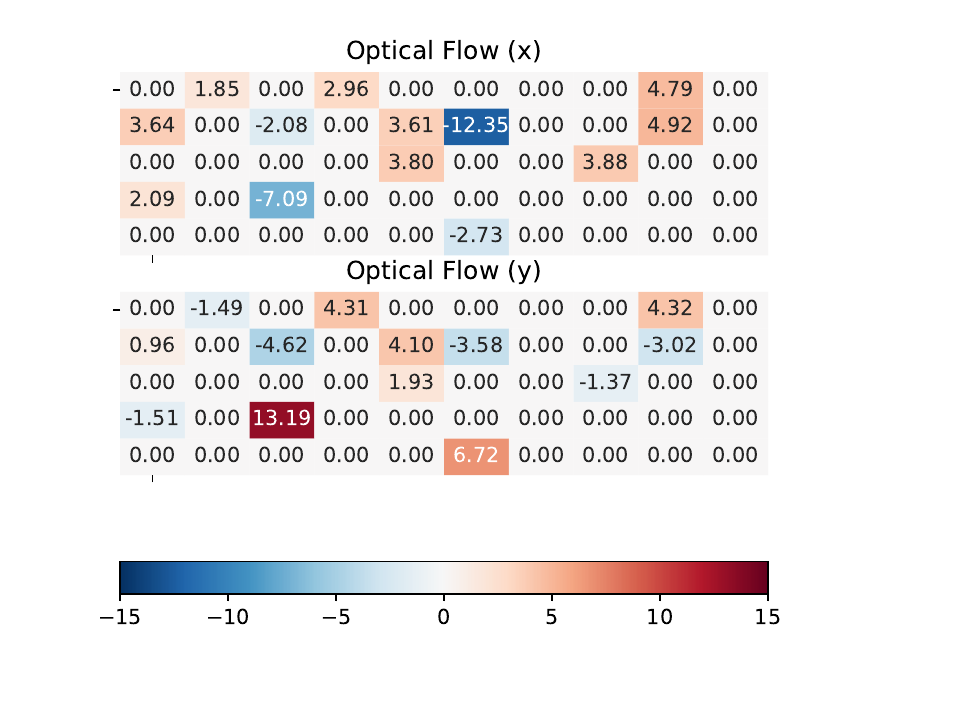}
  \caption{Optical flow representation when using regions of $10\times10$ pixels}
  \label{fig:Optical_flow}
\end{figure}

\subsubsection{Video pre-processing}
\label{ssec:videoprocessing}
First, we compute the optical flow with the method explained in~\cite{OpticalFlow}. To align the visual and audio features, we generated frames from the optical flow stream every $10$~ms. Over each frame. However, due to the nature of event-driven cameras, the number pixels that generate optical flow in each frame is different and therefore, the number of video features in each frame is different. To avoid this problem, we generate regions of same size across the $100\times50$ pixels.

For each region, we compute the mean of the $x$ component and $y$ component of optical flow and the event-rate, the number of events on each location at each frame. For example, with regions of $10\times10$ pixels we have a total of 50 regions and if we compute the event-rate and the mean of the $x$ and $y$ components, we have 150 video features. Fig.~\ref{fig:Optical_flow} shows an example of the $x$ component and $y$ component of the optical flow for a specific frame.

\subsubsection{Recurrent Neural Network}
\label{ssec:deepnetwork}
The RNN model consists of 5 stacked Bidirectional Long Short-Term Memory (BiLSTM) layers, with 250 neurons in each layer. The inputs of the model are the audio and visual features concatenated. The output of the network are an Ideal Amplitude Mask (IAM) and the loss function \(J_{mr}\):

\begin{equation}\label{iam}
    \mathbf{{p}}_t[f] = \dfrac{\mathbf{s}_t[f]}{\mathbf{y}_t[f]}
\end{equation}

\begin{equation}\label{loss}
	J_{mr} = \sum_{t=1}^T \sum_{f=1}^d (\mathbf{\hat{p}}_t[f] \cdot \mathbf{y}_t[f] - \mathbf{s}_t[f])^2\
\end{equation}
Where $p_t[f]$ is IAM, $s_t[f]$ is the target clean spectrogram, $y_t[f]$ is the noisy spectrogram and $\hat{p}_t[f]$ is the estimated IAM at each frequency bin $f \in [1,...,d]$.

We train the model using the Adam optimizer and 20\% of dropout to avoid overfitting. Each model is trained up to 500 epochs and early stopping is applied on the validation set to stop the training process.

%% file: results.tex
To measure the performance of each model, we use the well known source-to-distortion ratio (SDR) and PESQ~\cite{PESQ}, to quantify the separation of the target speech from the concurrent speech and the quality of cleaned speech (i.e. the speech enhancement measure), respectively.

\begin{figure}[!b]
  \includegraphics[width=\linewidth]{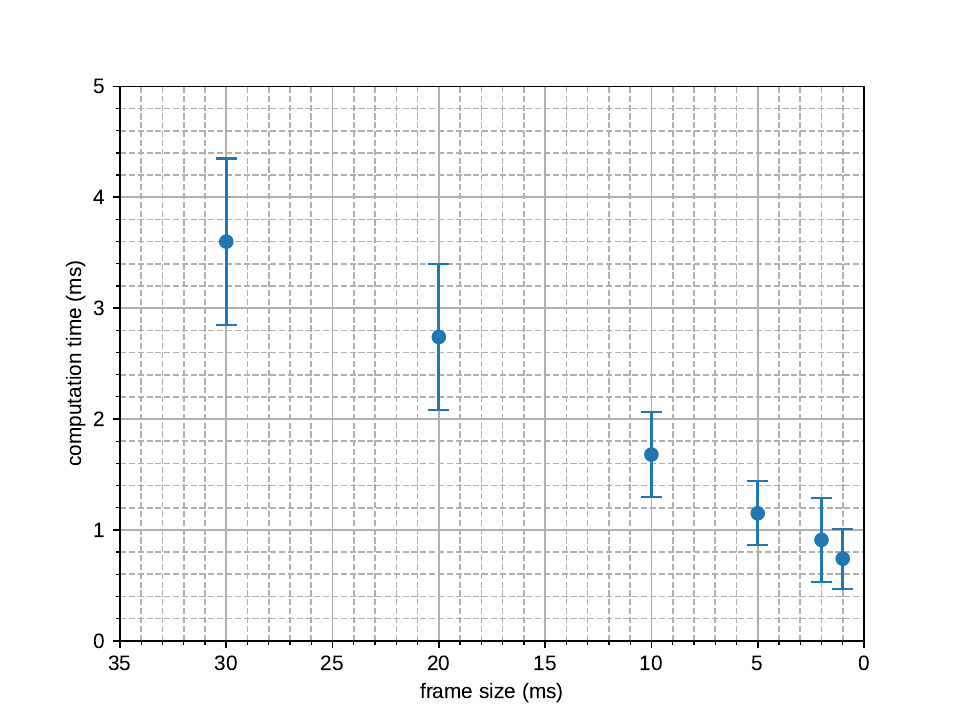}
  \caption{Computation time of optical-flow to extract visual features (Intel\textsuperscript{\textregistered} Core\texttrademark{} i7-7500U CPU @ 2.70GHz x 4)}
  \label{fig:Comp_time}
\end{figure}

Table~\ref{tab:results} shows the results from three different models. To train the first model we used 150 video features as input (concatenated with the audio features). These 150 features correspond to the $x$ and $y$ components of the optical-flow and the event-rate (the number of events) for each of the 50 regions ($10\times10$ pixels each region). The results are quite good, with higher than 7.0 SDR and on pare with the frame-based approach on PESQ performance. This shows that, although the original GRID dataset is frame-based, the event-based approach shows similar performance as the frame-based approach, despite the low quality and noisy events of the dataset used, that was not recorded live with the subjects, but obtained by recording a movie played back on a high resolution monitor.

In the next experiment we decrease the size of each region to $5\times5$ pixels in order to have more localized video features. However, the number of input features increases enormously. That is why we only used $x$ and $y$ components of the optical-flow not including the event rate like in the previous case(400 visual input features in total). Although the results are good (6.58 SDR and 2.59 PESQ), they are not close to those achieved with 150 input features. Besides, the training and inference time increases using 400 visual input features.

One of the drawbacks of BiLSTMs used in the previous experiments is that they need to pass all the features forward and backward before giving a prediction. That means that BiLSTM has higher latency than unidirectional RNN architectures. That is why we carried out one final experiment using LSTM instead of BiLSTM. However, the results show that the performance of deep LSTM is far from that yielded by the models with BiLSTM.

\begin{table}[]
\begin{tabular}{lll}
                                    & SDR           & PESQ          \\ \hline
Noisy signal                        & 0.21          & 1.94          \\
Frame-based approach~\cite{Morrone} & \textbf{7.37} & \textbf{2.65} \\
Event-based approach (150 features)  & 7.03          & \textbf{2.65} \\
Event-based approach (400 features) & 6.58          & 2.59          \\
Event-based approach (LSTM)         & 3.79          & 2.22          \\ \hline
\end{tabular}
\caption{GRID dataset results.}
\label{tab:results}
\end{table}

Finally, we compare the computation time of both approaches. Computing the face landmarks movements on frame-based approach for each video file (each video is 3 seconds long) the mean computation time using Dlib~\cite{Dlib} is 2.980 seconds with 0.825 standard deviation (SD).  On the other hand, for a event-based approach the mean is 1.126 seconds with 0.212 SD. The computation time of the event-based approach is almost three times less than the frame-based approach. The computation time of the event-based approach is divided as follows: 0.679 seconds computing the optical-flow and 0.447 seconds mapping optical-flow to the regions. In Figure~\ref{fig:Comp_time} the computation time of the optical-flow for different frame sizes is shown e.g. for a frame size of $10$~ms we accumulate events every $10$~ms to generate a frame and we compute the optical for the events in each frame. It can be seen that for all the cases the computation time of the optical-flow is lower than the frame size and there is enough time for mapping the optical-flow to the regions i.e. it is possible to extract the visual features before the next frame arrives without leaks. The hardware used to measure the computation time is and Intel\textsuperscript{\textregistered} Core\texttrademark{} i7-7500U CPU @ 2.70GHz x 4.

%% file: conclusions.tex
This work presents a RNN for target speaker audio extraction on multi-talker environment using event-driven camera for visual motion feature estimation. The system is trained on the GRID dataset. We show that, although this approach does not outperform the frame based approach in terms of quality of speech enhancement, the performance is almost on pair to the frame-based approach. We believe that improvements on the quality will be obtained when using data recorded directly with the EDC, as it will improve the spatial resolution and signal to noise ratio of the dataset.
At pair quality, the event-driven approach offers advantages in terms of computational cost and latency, that are critical for online, embedded applications.
The proposed method required the pre-processing of the frame-based dataset, using upscaling to 60 fps, for recording the event-driven dataset. However, this operation is only required once and won't be required in an online system where the visual signal is directly recorded by means of an EDC. The same linear interpolation operation needs to be always performed on the frame-based implementation to align video with audio features. The computation of the visual features depends on the scene, but is as low as 4ms, leading to a very low latency system implementation. To further reduce the latency of the output clean speech, we substituted the BiLSTM, that requires the passing of all the features forward and  backward, with an LSTM, however, the quality of the processed speech signal is far from being comparable to that of BiLSTM. Further work needs to address this problem.

To the best of our knowledge, this is the first work that uses event-driven cameras to address the target speaker extraction task. We showed that the method used to compute the optical flow to extract visual features is more efficient than the frame-based method used in~\cite{Morrone} and is better suited for embedded applications. Finally, this work shows that the $x$ and $y$ components of the optical flow from the lip region can be useful video features for target speaker audio extraction.
